# Modelling and Simulation of Scheduling Policies Implemented in Ethernet Switch by Using Coloured Petri Nets

B.Brahimi, C.Aubrun and E.Rondeau
Centre de Recherche en Automatique de Nancy
UMR-CNRS 7039
Faculté des Sciences, BP 239
54506 VANDOEUVRE LES NANCY Cedex

*Abstract*- The objective of this paper is to propose models enabling to study the behaviour of Ethernet switch for Networked Control Systems. Two scheduler policies are analyzed: the static priority and the WRR (Weighted Round Robin). The modelling work is based on Coloured Petri Nets. A temporal validation step based on the simulation of these modelling, shows that the obtained results are near to the expected behaviour of these scheduler policies.

## I. Introduction

Since the 1980s, a great deal of research has focused on the problem of distributed control over networks and the so-called Networked Control System (NCS) has developed. The source of this enthusiasm can be traced to the many advantages gained by eliminating the restrictions of traditional point-to-point control architectures. As an alternative to the point-to-point architecture, the NCS offers more flexibility for reconfiguration (e.g. to achieve fault-tolerance). Research on fault-tolerance aspects of NCS is at a very early stage of development and as such is a new requirement. Today most studies on fault-tolerance only include the effects of faults at either component or local controller levels. The autonomous fault-tolerant NCS is a distributed system involving fault diagnosis and control at various local to global levels of system embedding. Implementation of the subsequent concepts can be achieved by using the technologies of wireless networks, embedded systems, nomad components, electronics tags, etc...

Emphasis will be also placed on algorithms and procedures that facilitate the detection and isolation, at an early stage, of anomalies (variances or irregularities in the networks and/or in the system) and to switch to the fault-tolerant control strategy.

In particular, no significant theory in fault management and autonomous operating conditions exists, and only a few tools are available. Considerable efforts are still needed to make the range of theoretical results or methods in the control field applicable to networked systems. Integrated solution offering a synergy between communications, and computation and control, representing a new area of study for fault diagnosis and fault-tolerant control has to be developed.

In this context the aim of the NeCST (Networked Control System Tolerant to faults) project is to explore research opportunities in the direction of distributed control systems in order to enhance the performance of diagnostics and fault tolerant control systems. This leads to an improvement in the intensive use of NeCS technologies for the reactivity; autonomy and monitoring of large scale systems. The systems under consideration in the framework of this project can be considered as a distributed network of nodes operating under highly decentralised control, but unified in accomplishing complex system-wide goals. One of the key factors in designing such a complex system is that both the physical subsystem and the control part have to be designed together in an integrated manner.

The figure 1 shows the global approach of the NeCST project. The networked control system is decomposed in two parts: the network and the (industrial) application, which has to be studied in closed way.

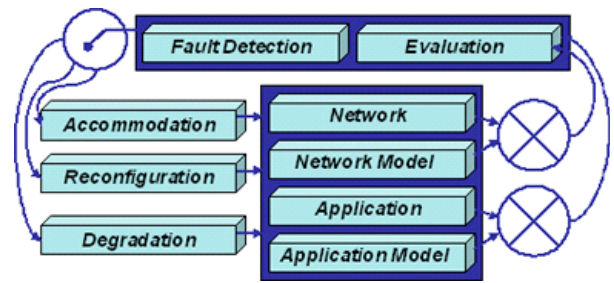

Fig. 1. Conceptual model of NeCST

The default detection is achieved in comparing the performance indicators provided by the real system (measured values) and by its model (estimated values). For network point of view, the communication information such as loss bandwidth, delay, and jitter can be monitored by using SNMP protocol and in a same time can be calculated from stochastic models or deterministic approaches. When the measured values and estimated values are different, it corresponds to one or more dysfunctions. The diagnostic has to enable to compensate the problem by using different mechanisms according to the importance of the fault:
- The accommodation is used in line and consists in only changing software parameters (traffic priority, scheduling policy,…)

- The reconfiguration temporarily stops the process for defining for example a new physical architecture in terms of network cabling, embedded components deployment,…
- The degradation modifies the application parameters, interrupts non-vital tasks in order to maintain the main functions of the process.

In the NeCST project, the communication system studied is the switched Ethernet architecture because it is more and more used in the Networked Control Systems. In this context, different works have been already investigated on the conceptual model of NeCST (figure 1).

Firstly, the network calculus theory is used to model the switched Ethernet architecture and to estimate the bounded end-to-end delays [3]. Secondly, the IEEE 1588 protocol (Precision Time Protocol) is implemented to synchronise the clocks of the different distributed devices. This enables to easily measure the delays. Thus when the measured delay is upper to the estimated bounded delay, it means that the network model does not correspond to the real network configuration and a fault is then detected (but not isolated).

In previous paper [1], accommodation mechanisms have been studied at application level. In this case, the evolution of network delays (estimated or measured) is compensated with the parameters of control models such as smith predictor model or robust control theory.

The problem is when the control model is not able to ensure the stability of the system due to a delay more important than the one originally estimated in the design of the NCS. In this case, a second level of compensating has to be considered and has to be applied on the communication system.

It is the goal of this paper. The objective is to analyse several compensation mechanisms implemented inside the network devices (Ethernet switch) in order to optimise the traffic according to the applicative importance of the messages. It corresponds to the adjustment between the quality of service offered by the network and the quality of control required by the application.

Some characteristics of Ethernet such as the nondeterministic access protocol and the propagation time of information in the network, raises questions on the capacity of these media to guarantee properties of determinism and reactivity of the control. Many works [2, 3, 4] showed that the use of switched Ethernet can be a solution to this problem. Protocols development [5] and new services offered by these technologies [6] gave place to many scientific works.

This paper addresses the problem of performances evaluation of the internal architecture of an Ethernet switch. When the Ethernet switch performances are disturbed by an external event such as instrument failure or additional load of the incoming traffic, transmission delay may increase.

The internal Ethernet switch architecture implements many complex mechanisms witch enable to ensure the exchanges of information between the nodes of application. Thus the communication behaviour representing all this complexity has to be modelled, verified and evaluated by using formal models.

The formalization of the communication architecture requires models presenting the following attributes [7]:

- Expression of parallelism, synchronization, the interaction, resource sharing;
- Expression of the temporal (and stochastic) characteristics associated to the mechanisms;
- Possibility of qualitative analysis (checking of the logic of the non-temporal and/or temporal mechanisms);
- Possibility of quantitative analysis (performance evaluation and/or reliability).

The approach presented in the paper is based on the well known Petri nets formalism. This formalism offers a framework well adapted and progressive for the representation and the analysis of the communications systems.

The model is obtained from the temporal Petri nets which are well adapted to the formulation of the problem studied is this paper. In adding, this work presents a generic and modular approach which requires the use of "colouring" and hierarchy of the model.

The temporal analysis of the hierarchical coloured temporal Petri net is obtained by simulation with CPNTools. This software developed by the Aarhus university to Denmark [8].

In the first section, the different components of the switch Ethernet are introduced. The second section is devoted to the explanation of the model structure. In this section, simulation results are presented for the following type of scheduler:
- Static Priority,
- Weighted Round Robin (WRR)

The first one tends to generate famine for the weak priority packets, and the second one has to solve this problem.

In the third section, the confrontation of the obtained models under a congestion situation of the switch is carried out. The results of simulation are discussed. Finally conclusions and perspectives are presented.

II. ETHERNET SWITCH MODELLING

*A. Introduction*

A Switch is a complex system which includes different mechanisms and technologies. [9] and [10] decompose the switching architecture in three main functional components:
- The queuing models refers to the buffering and the congestion mechanisms implemented in the switch,
- The switching algorithm implementation refers to the decision making process within the switch (how and where a switching decision is made),
- The switching fabric is the path that data take to move from one port to another.

There are different ways to build up the switch architecture with each of these components. In this deliverable, only one specific switching architecture is considered: Cisco Catalyst 2900 XL. Its modelling is shown at the figure 1 and uses the elementary components defined below:

One multiplexer and one queue to represent a switch using a shared memory (1&2),
- One demultiplexer to model the switching step (3),
- One demultiplexer for each output port (4)
- As much buffer as defined priorities (up to 8) for each output port (6),

- And one multiplexer for each output port defining the bandwidth used (7).

The figure 2 shows the model of a switch which manages two priorities on the frames.

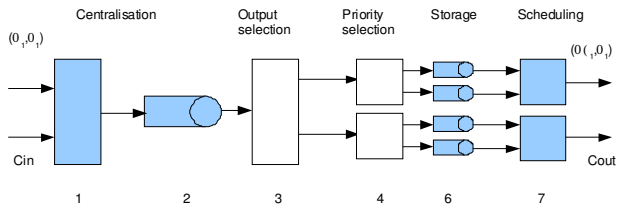

Fig. 2. IEEE 802.1 P/Q switch model

*B. Ethernet communication modelling by using Coloured Petri Nets*

Coloured Petri Nets (CPN), proposed by K.Jensen [11], is an extended version of classical Petri Net. In addition to places, transitions and tokens, the concepts of colours, guards and expressions are introduced so that computed data values can be carried by the tokens. This concept enables to introduce information (simple or complex) into the tokens. Moreover, the model of switch based on the CPN design tool is more compact.

A Coloured Petri net is graphical oriented language for design, specification, simulation and verification of systems. It is in particular well-suited for systems in which communication, synchronisation and resource sharing is crucial problem. CPN has an intuitive, graphical representation that is make its use easy.

Typical examples of application areas are communication protocols, distributed systems, automated production systems and others. Moreover, many works emerged to show the advantages of this kind of modelling [12, 13, 14, 15, 16].

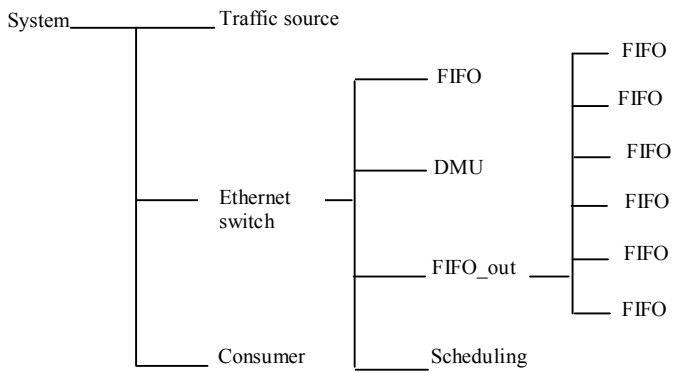

Fig. 3. Tree structure of the model

The model of the switch is built following a hierarchical and modular architecture. The structure of the model is represented in figure 3. Each node of the tree structure is a sub-model which corresponds to a substitution transition of the initial model.

The figure 4, shows the root of the hierarchical representation of the architecture of the model. The source of traffic whose activity is modelled by the transition traffic source generates packets to the switch by producing tokens at the Ptr1 place. The switch transmits the packets generated by traffic source towards the consumer via the places Ptr2 and Ptr2 '.

The places play the role of inputs/outputs for sub-models, the places Pbp1 and Pbp2 represents the availability of the output: when those are free, Pbp1 and Pbp2 are marked.

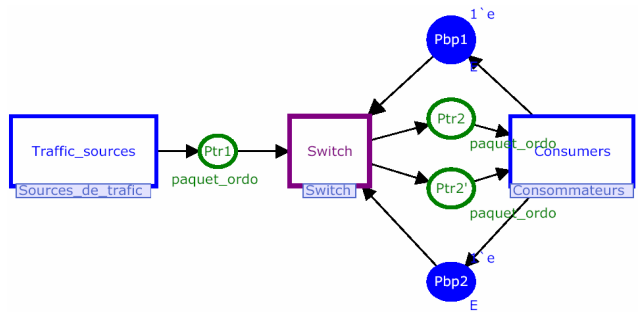

Fig. 4. Root level of the model

*C. Source modelling*

The colours associated to the tokens which represent the Ethernet packets are triplets: source equipment INP, destination equipment OUTP, and level of priority represented by PRIO.

We have three levels of priorities for our model, obviously, other attributes can be added. Each triplet supports temporal stamp.

Colorset packet = product INP*OUTP*PRIO timed.

The colour ord (see the label (i,g) in the figure 5)also called variable inherits the colorset packet. This variable models the packets which cross the different sub-models of figure 4. (traffic_source, swicth and consumers).

The packets are generated by the traffic source represented by the figure 5. It is a periodic source of period d (the period is equal to 5 in this figure). This period enables to model the sampling period of sensors and actuators.

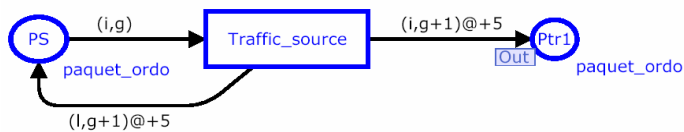

Fig. 5. Periodic source model

## D. Ethernet switch modelling

The packets thus generated by the traffic source to destination of the switch cross the different sub-modules constituting the Ethernet switch. The sub-model of the substitution transition of the switch is represented by the figure 6.

Firstly, the packets cross the FIFO queue. Secondly, they cross the demultiplexer. The goal is to route the packets to the output port (defined by attribute OUTP). And according to the priority which is associated to the packets (defined by attribute PRIO), the packets are stored in the FIFO queue (there is as many queue as of priorities). Finally the scheduling policy defines in the scheduler, processes the different packets.

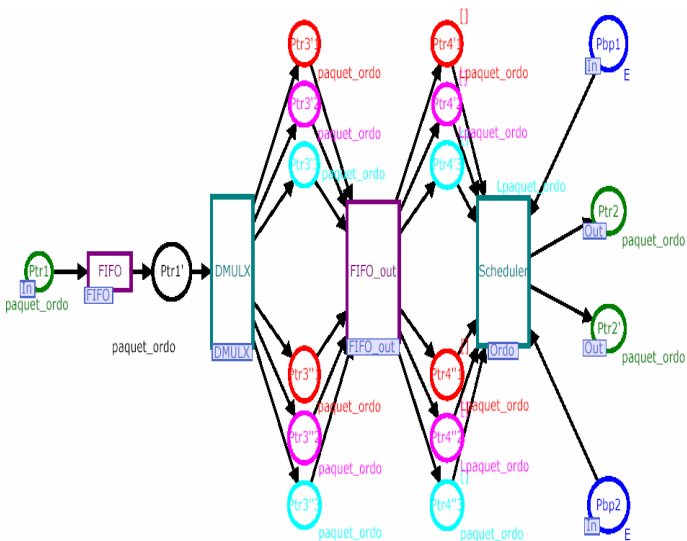

Fig. 6. Ethernet switch model

The output packets of the switch are consumed by a periodic consumer.

FIFO model is represented in the figure 7. This algorithm processes the packets in the order of their arrival. The main advantage of this algorithm is its simplicity of implementation. On the other hand, it does not make any distinction between flows, i.e. it offers only one level of service (it is not adapted to guarantee the quality of service).

To model a FIFO queue with CPNTools, we use the concatenation primitive .

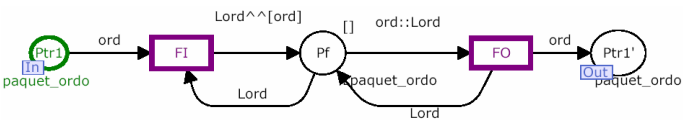

Fig. 7. FIFO queue model

The set of the demultiplexers is represented by the model described in the figure 8. Transition DMULX models the first demultiplexing according to the destination attribute associated with each packet,. In the figure 8, two destinations are modelled O1, O2 (which correspond to output 1 and 2, respectively). Technically, this demultiplexing is defined by crossing conditions (or not) associated to the output arcs of the transition DMULX.

The function of the transitions DMULX1 and DMULX2 is to route the packets according to their priorities. In this study, three priority levels are defined: H: high level of priority; M: average priority and b: low priority. In the same manner that the transition DMULX, the routing is carried out by conditions associated to the output arcs of the two transitions.

Then, the FIFO queues which represent the different priority levels, receive the packets coming from the demultiplexers (DMULX1 and DMULX2).

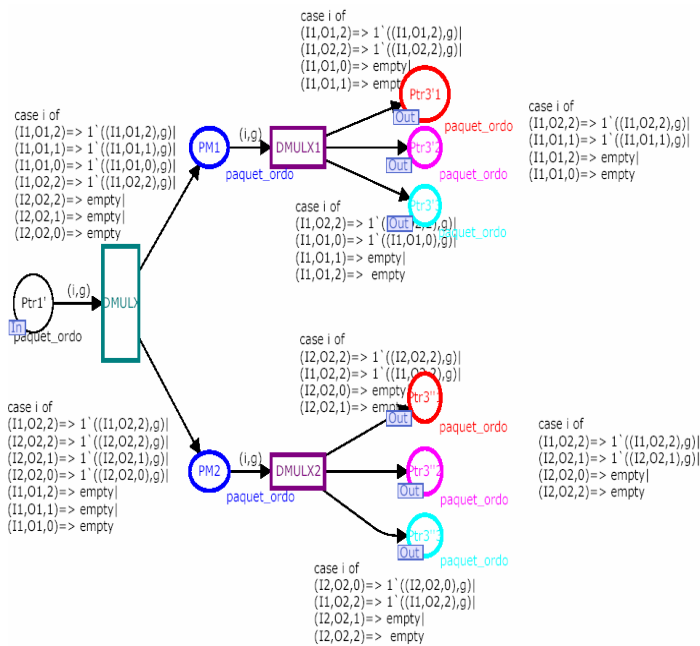

Fig. 8. Demultiplexer model

Finally, the packets stored in the FIFO queues are processed according to their priority, and also in taking into account the selected scheduling policies. In the next section, we model the static priority policy and Weighted Round Robin (WRR).

## E. Static priority modelling

The model of the figure 9 describes the behaviour of a static priority scheduling. The type of packets is classified in three groups:
- Packets of flows which have a priority higher than the other packets. These packets are called packets H and are modelled by tokens in the place Ptr4' 1.
- The packets which belong to flows of priority lower than the latter one. They are modelled by tokens in the place Ptr4' 2.
- The packets which belong to flows having the lowest priority.

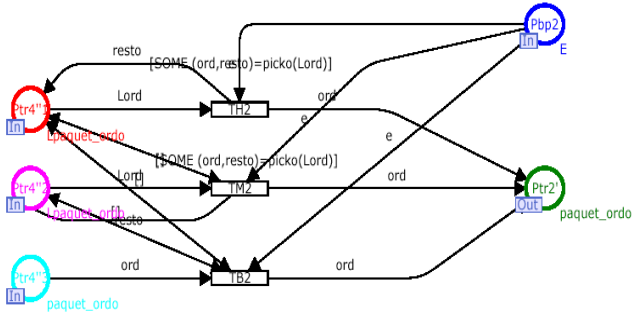

Fig. 9. Priority static scheduler model

A packet with lowest priorities can be delayed by the other packets due to the non-pre-emptive characteristics of this kind of packet scheduling algorithms. This situation occurs when a higher priority packet arrives during the transmission of the packet with a low priority. Even if the packet has a higher priority it has to wait until the low priority packet is fully transmitted.

The interest of the static priority algorithms is that it is easy to implement. But it manages the bandwidth in iniquity way. The consequence is the low priority packets could be not transmitted (famine problem) and could generated perturbations on the controlled systems. To eliminate the iniquity situation, the Fair Queuing algorithm is studied in the next section.

*F. Weighted Round Robin modelling*

The Fair Queuing (FQ) algorithm enables to share the bandwidth equitably between several flows. A weight is associated to each flow, defining the amount of bandwidth which must be allocated to it. The implementation of such algorithm is very complex.

We propose to use a similar algorithm called the Round-Robin (RR) algorithms and easier to implement.

RR Algorithms assign a queue to each flow. In a cycle way, the server visits the queues recurrently to process the packets on stand-by as service.

The duration of the service is fixed by the service policy defined by RR algorithm. At the end of service allocated time, the server visits the following queue considered in the cycle. And, the same procedure starts again.

The most known RR algorithms in practice are:
- Weighted Round Robin (WRR)
- Deficit Round Robin (DRR).

The WRR has the capability to deal with the problem of iniquity which is the main problem addressed in this paper.

WRR assigns a weight of Wi to each queue i which fixes the service quantity. For each visit of queue i, the server transmits the packets which are stored until either the queue is empty or the number of the served packets reaches wi.

The figure 10 represents WRR algorithm for three priority levels.

When the server visits the queue represented by Ptr4' 1 (place of reception of the high priority packets), the place of the type "E" get marked. For each visit of queue the number of packet transmission is limited to wi. The place of the "wfi" type counts the number of packets which have been served during the current visit of the server. This place is put at wi on each arrival of the server to the queue. Its marking is then decremented when each packet is transmitted (i.e. following the transition TH1 firing). When the server leaves the queue this place is set to zero by the transition firing which has a guard of [Lw<>nil]

The server leaves the queue when the place Ptr4' 1 becomes empty, and/or when Wf (allocated time) is reached.

When the service of this queue is finished, the server processes the next queue in the cycle (represented by the place Ptr4' 1 which receives the mean priority packets). If this place is empty, the server waits for new incoming packets in the buffer (Ptr4' 1). This step is necessary when the queues are empty to avoid that the transitions are always fired. In this case the graph is not bounded.

The PN is as many duplicated once as there is various priority levels. The advantage the coloured Petri nets is to reduce the size of the initial PN by folding. The figure 10 represents the CPN without folding.

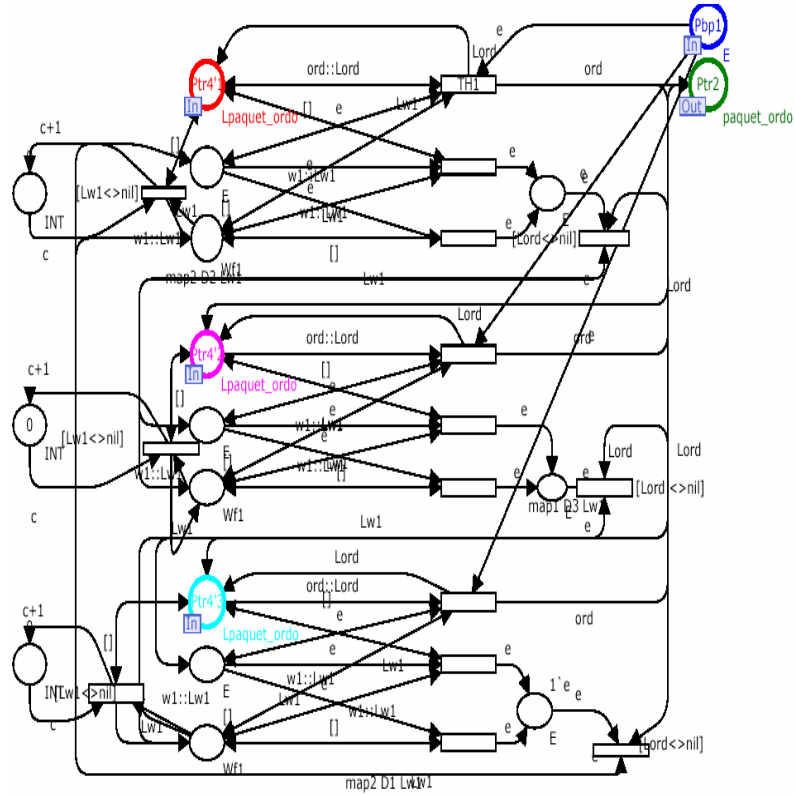

Fig. 10. WRR model

*G. Consumer modelling*

The packets are transmitted under the condition that the transmission line is free (the places Pbp1, Pbp2 must be marked).Then the packets are consumed by periodic consumers as shown in the figure 11.

As we consider an Ethernet switch with two output ports, two consumers are modelled. Place C1 represents a counter of the consumed packets for the different priorities. The counter is incremented if the condition associated to the input arc of this place is verified.

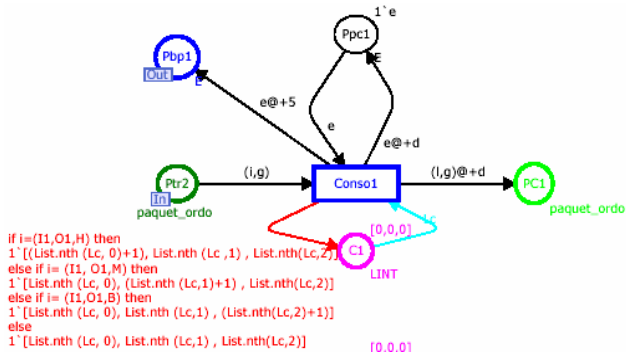

Fig. 11. Periodic consumer model

### H. Conclusion

The general model of an Ethernet switch is obtained by assembling the different sub-models described in this section.

For example, the figure 12 (see the last page) describes a hierarchical model of an Ethernet switch implementing the static priority scheduler.

## II. SIMULATION OF ETHERNET SWITCH MODELS

### A. Introduction

The objective of this section is to test 2-ports Ethernet switch model and to analyse the behaviour of the two schedulers: static priority and WRR. We have defined a scenario which collects six periodic source traffic and six periodic consumers.

In order to simulate the model (see the figure 12), the six sources of periodic traffic (period d = 5 time unit) are folded in only one. Three periodic consumers (period d = 5 t.u) are associated with the first output port of switch. And three other periodic consumers with the same period are attached to the second one. For each output port the three periodic consumers are folded in only one.

### B. Static priority scheduler simulation

The simulation of the model for 10.000 simulation steps and 565 t.u, give the results shown in the figure 13. it shows that 98,24 % of the generated packets of high priority are consumed, and only 0,87% of the mean priority packets are consumed and no packet of low priority is consumed. The end to end delay of high priority packets is 10 t.u and the standard deviation is nil.

We can carry out a quantitative analysis according to the results obtained. The static priority scheduling processes immediately the high priority packet, at the expense of other lower priorities packets. The advantage of the static priority algorithm is to be easy to implement.

Nevertheless, it shares the bandwidth in iniquity way. The static priority algorithm generates famine for the consumers which are waiting for the mean and low priority packets.

The consequence of this problem could destabilize the process.

Then, this model is simulated in the case where the switch is in a congestion state (figure 14). For this, for each source 2 packets with a period equal to 5 t.u are added on the previous scenario.

The obtained results show that the packets are consumed with a delay (for example: we can observe in the figure 14 a delay upper to 200 t.u). In the worst case, some packets are not consumed such as packets 35, 36… The average value of the end-to-end delay is of 50,38 t.u. The standard deviation is equal to 34,24 t.u and generates jitters on the network. The packets with other priority are not served.

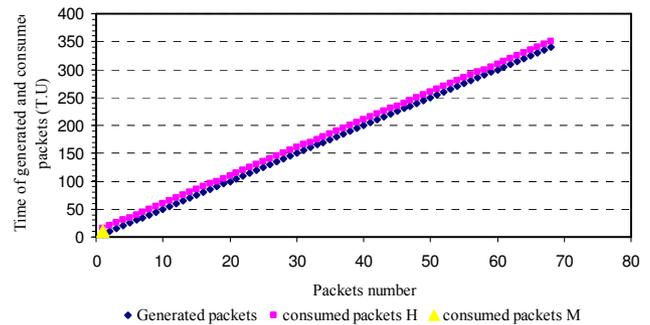

Fig. 13. Static priority scheduler simulation without congestion

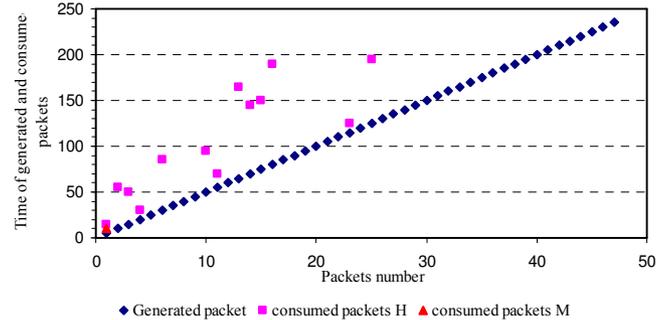

Fig. 14. Static priority scheduler simulation with congestion

### C. WRR simulation

The model was simulated for 10000 simulation steps and 565 t.u. At first we analyse the previous scenario without congestion in taking into account the WRR algorithm. With the same parameters attributed to the traffic sources and to the consumers. WRR algorithm has three queues corresponding to each priority. A weight is attached to a priority as follows: W1 (60%), W2 (30%), W3 (10%). The percentages represent the weights assigned to the high, mean, low priority packets, respectively.

The figure 15 represents the results obtained with this method.

The results show that 59,65 % of packets generated with high priority are consumed, and 28,94% of the generated of mean priority packet are consumed and 10,52% for the low priority packets. The end to end delay of the high priority packet is 11,70 t.u and the standard deviation is 2,64 t.u. For the mean priority packet, the end to end delay is 10 t.u and a standard deviation is nil. Finally, for the low priority packets the end to end delay is 9,44 t.u and the standard deviation is 1,66 t.u.

The simulation results clearly show equity (fairness) of consumption between the various packets, according to the predefined weights. The main advantage of this algorithm is to avoid the famine phenomenon. The interest of WRR algorithm in the context of NCS is to be able to adjust the weights in considering the application constraints to avoid process instability.

Secondly, we analyse the previous scenario with congestion.
The figure 16 shows the switch is overloaded.

Moreover, only 28,98% of the generated with high priority packets are consumed, 15,94% of mean priority packets and 5,79% of low priority packet are consumed. The end to end delays are 9,54 t.u, 10 t.u and 10 t.u, respectively. The standard deviation are 1,50 t.u, 0 t.u and 0 t.u for the high, mean and low priority packets, respectively.

The interests of WRR algorithm in the congestion state are to minimise the end to end delay and the standard deviation in regard to the results obtained from the static priority algorithm.

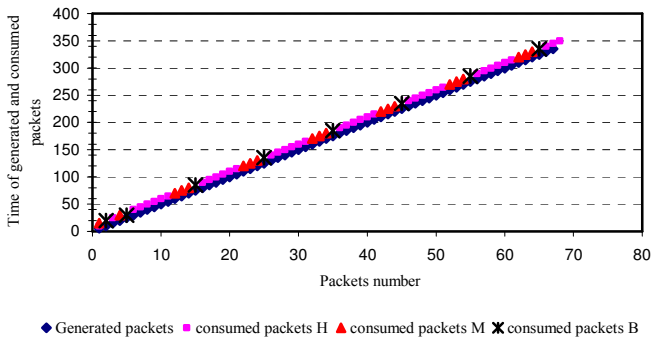

Fig. 15. WRR scheduler simulation without congestion

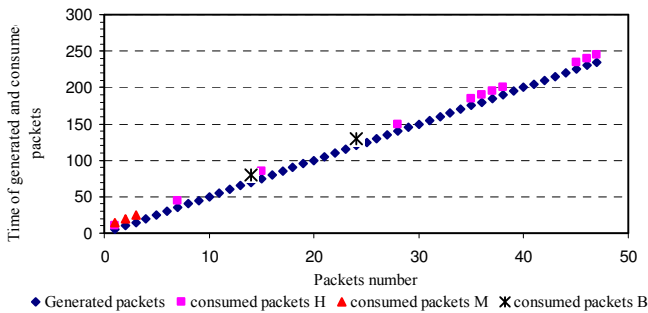

Fig. 16. WRR simulation with congestion

## III. Conclusion

This paper proposes a general model to represent the behaviour of an Ethernet switch by using CPN. Two schedulers have been modelled and evaluated showing the interest to implement a WRR policy which solves the iniquity problem.

In the future work, the goal of our research is to propose algorithms which enable to adjust the weights defined in the WRR algorithm in considering the time constraints of process.

The benefit of this approach is to avoid instability situation in combining the quality of service offered by the network and the quality of control defined by the application.

These algorithms will be embedded in each network devices to be able to compensate dynamically the impact of the delay on the networked control system.

Finally, we are studying other scheduling policies such as the Earliest Deadline First (EDF) which is interesting to dynamically control the network [17].

ACKNOWLEDGMENT

The authors wish to acknowledge the funding support for this research under the European Union 6th Framework Program contract n° IST – 2004-004303 Network Controles Systems Tolerant to faults (NeCST).

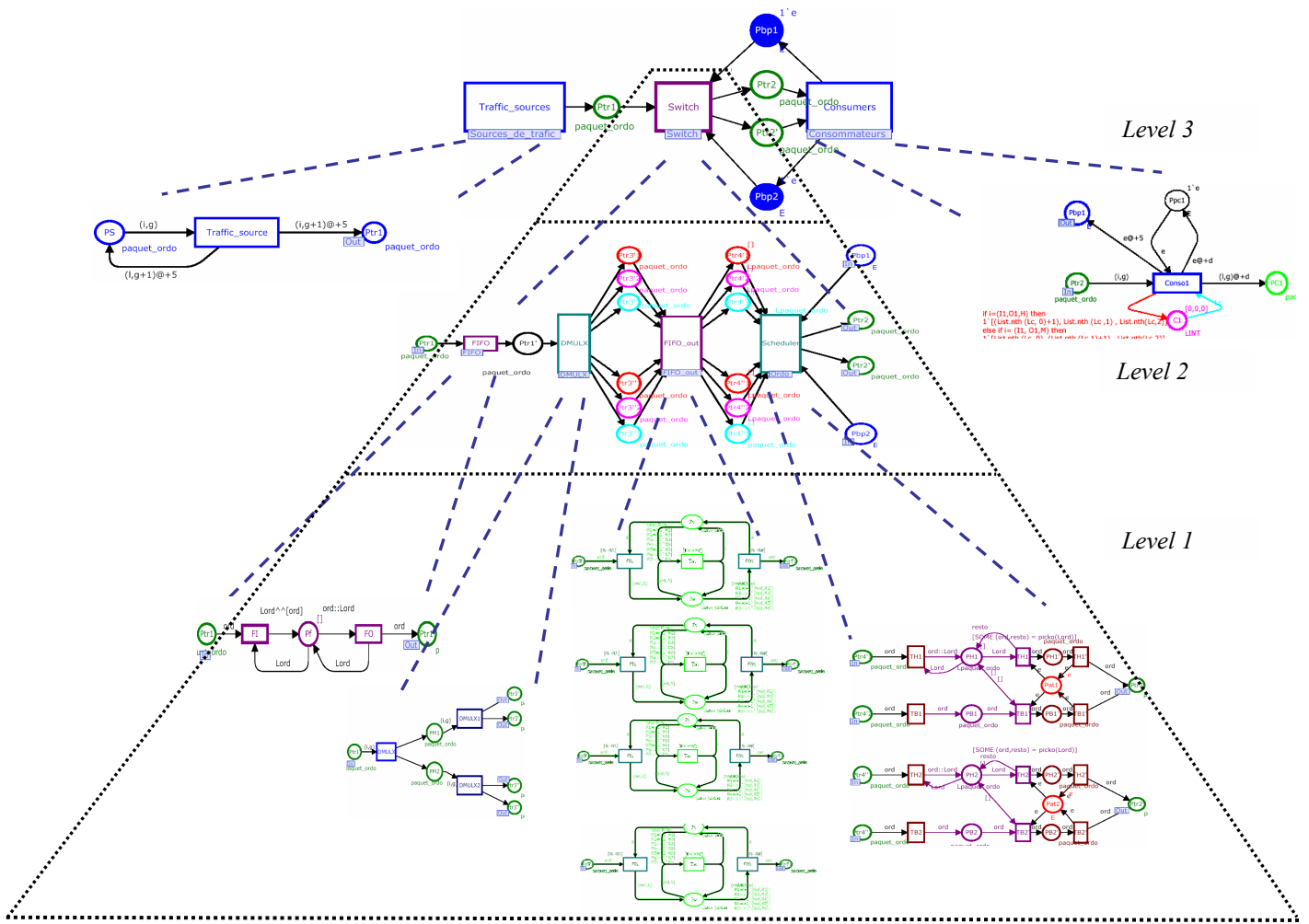

Fig. 12. Hierarchical model of Ethernet switch